\documentclass[useAMS,usenatbib]{mn2e}
\usepackage{aas_macros,graphicx,times,multirow,amsmath,bbold}
\usepackage[]{color,hyperref}
\def\src{1RXS J1708}

\title[Magnetars X-ray polarization]{Probing magnetars magnetosphere through X-ray polarization measurements}
\author[R. Taverna et al.]{R.~Taverna$^{1}$\thanks{E-mail:
\href{mailto:taverna@pd.infn.it}{taverna@pd.infn.it}},
F.~Muleri$^{2}$, R.~Turolla$^{1,3}$, P.~Soffitta$^{2}$,
S.~Fabiani$^{2}$, L.~Nobili$^{1}$
\smallskip\\
$^1$Dipartimento di Fisica e Astronomia, Universit\`a di Padova, Via Marzolo 8, I-35131 Padova, Italy\\
$^2$INAF-IASF Roma, Via del Fosso del Cavaliere 100, I-00133 Roma, Italy\\
$^3$Mullard Space Science Laboratory, University College London, Holmbury St. Mary, Dorking, Surrey, RH5 6NT, UK
}
\date{Accepted \ldots. Received \ldots; in
original form \ldots} \pagerange{\pageref{firstpage}--\pageref{lastpage}} \pubyear{2013}

\def\LaTeX{L\kern-.36em\raise.3ex\hbox{a}\kern-.15em
    T\kern-.1667em\lower.7ex\hbox{E}\kern-.125emX}

\def\flux {\mbox{erg cm$^{-2}$ s$^{-1}$}}
\def\lum {\mbox{erg s$^{-1}$}}

\begin{document}

\label{firstpage}
\maketitle
\begin{abstract}
The study of magnetars is of particular relevance since these objects are the only laboratories where the physics in ultra-strong magnetic fields can be directly tested. Until now, spectroscopic and timing measurements at X-ray energies in soft $\gamma$-repeaters (SGRs) and anomalous X-ray pulsar (AXPs) have been the main source of information about the physical properties of a magnetar and of its magnetosphere. Spectral fitting in the $\sim 0.5$--10 keV range allowed to validate the ``twisted magnetosphere'' model, probing the structure of the external field and estimating the density and velocity of the
magnetospheric currents. Spectroscopy alone, however, may fail in disambiguating the two key parameters governing magnetospheric scattering (the charge velocity and the twist angle) and is quite insensitive to the source geometry. X-ray polarimetry, on the other hand, can provide a quantum leap in the field by adding two extra observables, the linear polarization degree and the polarization angle. Using the bright AXP 1RXS J170849.0-400910 as a template, we show that phase-resolved polarimetric measurements can unambiguously determine the model parameters, even with a small X-ray polarimetry mission carrying modern photoelectric detectors and existing X-ray optics. We also show that polarimetric measurements can pinpoint vacuum polarization effects and thus provide an indirect evidence for ultra-strong magnetic fields.
\end{abstract}

\section{Introduction}
\label{intro}

Soft gamma repeaters (SGRs) and Anomalous X-ray pulsars (AXPs)
form together a class of neutron star X-ray sources characterized
by a number of peculiar properties: emission of short ($\approx
0.1$--1 s), energetic ($\approx 10^{38}$-- $10^{41}\, \lum$) X-ray
bursts, occurrence of outbursts, i.e. sudden enhancements (up to a
factor $\approx 1000$) of the persistent flux of duration $\approx
1$ yr, quite long spin periods ($\sim 2$--$12$ s) and large (as
compared to ordinary radio pulsars, PSRs) spin-down rates
($\approx 10^{-13}$-- $10^{-10}\, \mathrm{ss}^{-1}$). Three SGRs
have been observed to emit also giant flares, hyper-energetic
events in which a luminosity $\approx 10^{44}$-- $10^{47}\, \lum$
at the peak was released over a timescale of a few hundred seconds
\cite[e.g.][for reviews]{mereg08, reaesp11,turesp13}.

SGRs/AXPs are convincingly associated with an isolated neutron
star (NS), do not appear to be powered by rotational energy losses
and are usually radio-silent, at variance with PSRs, although the
existence of a continuum of properties across the two groups
starts to emerge \cite[e.g.][]{reaetal10,reaetal12}. There are by
now several independent indications that SGRs/AXPs are magnetars,
i.e. their activity is sustained by the magnetic energy stored in
the (internal) field of an ultra-magnetized NS. Very recently,
\cite{tiengoetal13} reported the discovery of a proton cyclotron
feature in the X-ray spectrum of the ``low-field'' magnetar SGR
0418+5729 \cite[][]{reaetal10}, showing that ultra-strong
(localized) magnetic structures with $B\approx 10^{15}\, \mathrm
G$ are present near the surface of this neutron star.

The magnetar model has been quite successful in explaining the
overall properties of SGRs/AXPs , both concerning their bursting
and persistent emission. The latter is characterized by a
luminosity $L_{\mathrm X}\approx 10^{31}$-- $10^{36}\, \lum$ in
the $\sim 0.2$--$10$ keV range with a spectral distribution which can
be approximated by the superposition of a blackbody component at
$kT\sim 0.5\, \mathrm{keV}$ and a high-energy, power-law tail,
with photon index $\Gamma\approx 2$--4\footnote{Some transient
magnetars exhibit a nearly thermal spectrum, modelled by one, or
more, blackbody component(s) \cite[e.g.][]{reaesp11}.}. According
to the ``twisted magnetosphere'' model (Thompson, Lyutikov \& Kulkarni 2002, \citeauthor{TLK} hereafter), the
external magnetic field of a magnetar acquires a toroidal
component (the ``twist''), as a consequence of the crustal
deformations induced by internal magnetic stresses. Twisted fields
are non-potential and require supporting currents to flow along
the closed field lines. The density of charged particles (mainly
$e^\pm$) is high enough to make the magnetosphere thick to
resonant cyclotron scattering (RCS). Thermal photons emitted by
the cooling star surface undergo (multiple) Compton scatterings
onto the moving charges and fill the non-thermal tail of the
spectrum. Detailed radiative transfer calculations based on Monte Carlo methods
confirmed this picture (\citeauthor[][]{ft07} 2007; Nobili, Turolla \& Zane 2008a,b, see also \citeauthor{lg06} 2006).

Although current RCS models rely on a number of simplifying assumptions, mainly a ``globally twisted'' magnetosphere and
rather ad hoc space and velocity distributions of the scattering charges (see \citeauthor{ft07} 2007; Nobili, Turolla \& Zane 2008a, \citeauthor{NTZ1} in the following),
their systematic application to fit SGRs/AXPs X-ray spectra has been largely successful, allowing to validate the
``twisted magnetosphere'' scenario and to estimate some of the magnetospheric parameters \cite[][]{reaetal08,zrtn09}.
Theoretical work to overcome some of these limitations is under way \cite[e.g. by considering non-global
twists,][and calculating currents from first principles, \citealt{bt07,belo13}]{belo09,pav09}, but a fully consistent
picture of the interaction of radiation with the flowing charges in a magnetar magnetosphere is still to come.

Comparison of RCS models with X-ray spectral data is not bound, in any case, to provide complete information. Due to
an
inherent degeneracy in the RCS model parameters, in fact, spectral fitting alone may 
be insufficient to unequivocally determine both the twist angle and the charge velocity. Moreover, computed spectra
are rather insensitive to the source geometry, although in principle they
do depend on the angles that the line of sight and the magnetic axis make with the star rotation axis
(\citeauthor{NTZ1}; \citeauthor{zrtn09} 2009). While the
degeneracy may be removed by performing a simultaneous fit of both the (phase-averaged) spectrum and the pulse profile
\cite[][]{alba10}, polarization measurements at X-ray energies can disclose an entirely new approach to the
determination of the physical parameters in magnetar magnetospheres.

Radiation traversing a strongly magnetized vacuum, such as that
around a neutron star, propagates into two normal modes, the
ordinary (O) and extraordinary (X) mode \cite[e.g][]{Harding+Lai}.
X-ray radiation from a magnetar is expected to be polarized for
essentially three reasons: i) primary, thermal photons, coming from the star surface, can be
intrinsically polarized, because emission favors one of the modes
with respect to the other; ii) scattering can switch the photon
polarization state; and iii) once the scattering depth drops, the
polarization vector changes as the photon travels in the
magnetosphere outside the ``adiabatic region'' \cite[the so called
``vacuum polarization''][see also \citealt{Harding+Lai}]{hs00,hs02}.

Fern\'{a}ndez \& Davis (2011) (\citeauthor{Fernandez+Davis}
hereafter) have presented a comprehensive study of the
polarization properties of magnetar radiation in the X-ray band,
with a view to a polarimeter which was to fly on the (now
cancelled) mission GEMS. In this paper we complement the work by
Fern\'{a}ndez \& Davis and reexamine by means of detailed Monte
Carlo simulations how X-ray polarization measurements performed by
next-generation instruments, the XIPE polarimeter in particular
\citep{Soffitta2013}, will allow to exploit magnetars as
laboratory for fundamental physics and give a new and unique
insight into their magnetospheric environment.

\section{The model} \label{sec:theoreticalmodel}

In this section we discuss the physical bases for our calculations of the polarization properties of magnetars X-ray
emission.

\subsection{Magnetospheric geometry and RCS} \label{subsec:magnetosphere}

The super-strong (up to $10^{16}\, \mathrm G$) internal magnetic
field of magnetars is believed to be highly wound up, with
toroidal and poloidal components roughly of the same order. The
huge magnetic stresses acting on the star crust induce
deformations/fractures, allowing some of the magnetic helicity to
be transferred to the external field and powering SGRs/AXPs
activity (see \citeauthor{TLK}; \citeauthor{pp11} 2011). As a consequence, an external
toroidal component, $B_\phi$, builds up, twisting the
magnetosphere. Twists are likely localized into bundles of field
lines with footpoints anchored in the regions which underwent a
relative displacement \cite[][]{belo09}. In the following,
however, we will use the simplified model originally introduced by
\citeauthor{TLK} (see also \citeauthor{ft07} 2007, \citeauthor{NTZ1}), who
considered an axisymmetric, globally sheared dipole field which in polar components is
\begin{equation}\label{bgen}
  {\boldsymbol{B}} = \frac{B_{\mathrm{p}}}{2} \left(\frac{r}{R_{\mathrm{NS}}}\right)^{-p-2}
  \left[-f', \frac{pf}{\sin\theta},
    \sqrt{\frac{C(p)\ p}{p+1}} \frac{f^{1+1/p}}{\sin\theta}\right]
\end{equation}
where $R_{\mathrm{NS}}$ is the NS radius, $B_{\mathrm{p}}$ is the value of the magnetic field at the pole,
$r$ and $\theta$ are the radial coordinate and magnetic colatitude, respectively, and a prime denotes
derivation with respect to $\cos\theta$.
The function $f=f(\cos\theta)$ satisfies the Grad-Shafranov equation and is easily computed numerically,
together with $C(p)$, once the value of the radial index $p$ is fixed (see \citeauthor{TLK}; \citeauthor{pav09} 2009).
The amount of shear is usually measured through the twist
angle, defined as

\begin{equation} \label{eqn:Df}
\begin{aligned}
&\Delta\phi_{\mathrm{N-S}} &=&\ \ \ \ \  \lim_{\theta\rightarrow 0} \int_{\theta}^{\pi/2}
\frac{B_{\phi}}{B_{\theta}\sin{\theta}}\mathrm{d}\theta \\
&\ &=&\ \ \ \ \ \left[\frac{C(p)}{p\ (1+p)}\right]^{1/2}
 \lim_{\theta\rightarrow 0}  \int_{\theta}^{\pi/2} \frac{f^{1/p}}{\sin\theta} \mathrm{d}\theta \,.
\end{aligned}
\end{equation}

While in a perfectly dipolar magnetosphere charged particles can flow only along the open magnetic field lines
\cite[the Goldreich-Julian currents,][]{Goldreich+Julian}, in the magnetar case, where the external
magnetic field is
non-potential ($\boldsymbol{\nabla}\times\boldsymbol{B}\neq 0$), currents must also circulate along the closed field lines.
The details of these currents are still not completely explored, but they appear to be dominated by pairs, created by photons as they interact with primary electrons in the ultra-strong field \cite[][]{bt07}. In the simplest case in which
the charge carriers are electrons and ions (unidirectional flow), the spatial
density of the magnetospheric particles follows from the requirement that the current density is $\boldsymbol{j}=c{\boldsymbol{\nabla}}\times {\boldsymbol{B}}/4\pi$,
\begin{equation}\label{magcurr}
n_{\mathrm{e}} = \frac{ p + 1 } {4 \pi e} \left ( \frac{B_\phi}{B_\theta} \right ) \frac {B }{r\vert  \langle \beta \rangle\vert } \, ,
\end{equation}
where $\langle\beta\rangle$ is the average charge velocity (in units of $c$, \citeauthor{TLK,NTZ1}). Electrons are assumed
to have a 1-D (relativistic) Maxwellian distribution at $T=T_{\mathrm{el}}$, superimposed to the bulk motion along the field lines
(\citeauthor{NTZ1}).

As shown by \citeauthor{TLK}, the electron density implied by equation (\ref{magcurr}) is large enough to make
the magnetosphere thick to resonant (electron) cyclotron scattering, so a photon of energy $\hbar\omega$ will scatter when
the condition
\begin{equation}\label{omegad}
\hbar\omega=\frac{\hbar\omega_{\mathrm{B}}}{\gamma(1-\beta\cos\theta_\mathrm{bk})}
\end{equation}
is met; here $\hbar\omega_{\mathrm{B}}$ is the electron rest-frame cyclotron energy and $\theta_{\mathrm{bk}}$ is the
angle between the
incident
photon direction and the particle velocity, $\beta$ ($\gamma$ is the Lorentz factor). In SGRs/AXPs, thermal photons
are emitted from the star surface with a typical energy
$E\sim 1$ keV and scatter at a few star radii, where the magnetic field has decayed to  $B\approx 10^{11}\, {\rm G}$.

\subsection{Polarization of radiation} \label{subsec:polarization}

In the presence of a strong  magnetic field, the vacuum around the star behaves as a birefringent medium,
in which photons propagate in two normal modes of polarization: the ordinary mode (O-mode), with
the electric field in the $\boldsymbol{\hat{k}}-\boldsymbol{B}$ plane, and the extraordinary mode
(X-mode) with the electric field perpendicular to this plane (here $\boldsymbol{\hat{k}}$ is the photon direction,
e.g. \citealt{Harding+Lai}). Thermal photons
coming from the stellar surface can be polarized either in the O- or in the X-mode. Nevertheless, at the surface
it is $\hbar\omega \ll \hbar\omega_{\mathrm{B}}$, and in the  hypothesis
that the photon energy is far enough away from the ion cyclotron energy, the opacity for X-mode photons, which goes as
$\kappa_{\mathrm{X}}\sim \kappa_{\mathrm{O}}(\omega/\omega_{\mathrm{B}})^2$, is much less than that for the O-mode
\cite[e.g.][]{Harding+Lai,Laietal}. So, under
these conditions, the seed thermal radiation is likely to be mostly polarized in the X-mode. Simulations presented in the rest of the paper conform to this picture, although the polarization fraction of thermal radiation is not completely assessed as yet. In this respect \citet{bt07} noted that thermal emission from the surface regions heated by the returning currents should preferentially occur in the O-mode.

Due to RCS, photons can change their polarization state. By
evaluating the expression for the RCS cross sections, it turns out
that an O-mode photon is more likely scattered into the X-mode,
while an X-mode one has a greater probability to retain its
initial polarization state (e.g. \citeauthor{NTZ1}). More specifically,  the (total) scattering cross
sections are
\begin{equation}
\sigma_{\mathrm{O}-\mathrm{O}}=\frac{1}{3}\sigma_{\mathrm{O}-\mathrm{X}}, \quad
\sigma_{\mathrm{X}-\mathrm{X}}=3\sigma_{\mathrm{X}-\mathrm{O}}\,,
\end{equation}
where the first subscript refers to the incident and the second to the scattered photon polarization mode.


The strong magnetic field has itself a direct effect on the
polarization of radiation travelling in the magnetosphere.
Photons, in fact, can convert into virtual $e^\pm$ pairs
because of vacuum polarization, as predicted by QED. The external
magnetic field modifies the vacuum dielectric and magnetic permeability tensors according to
\begin{equation} \label{eqn:epsilonmu}
\begin{aligned}
&{\boldsymbol{\varepsilon}} & = &\ \ \ \ \ (1+a)\mathbb{1}+q{\boldsymbol{\hat{B}}}{\boldsymbol{\hat{B}}} \\
&{\boldsymbol{\bar{\mu}}} & = &\ \ \ \ \
(1+a)\mathbb{1}+m{\boldsymbol{\hat{B}}}{\boldsymbol{\hat{B}}}
\end{aligned}
\end{equation}
where ${\boldsymbol{\bar{\mu}}}$ is the inverse of the magnetic
permeability tensor, $a$, $q$ and $m$ are functions of the
magnetic field intensity and ${\boldsymbol{\hat{B}}}$ is the unit vector along
the magnetic field \cite[ for the case at hand only the low-field approximation for $a$, $m$ 
and $q$ is required; 
see 
e.g.][for more
details and the Appendix]{Harding+Lai}. The plasma contributions to the dielectric
and magnetic permeability tensors are negligible compared to the
QED ones up to $\sim 3000\, R_{\mathrm{NS}}$ under the typical
conditions of a magnetar magnetosphere (e.g. \citeauthor{Fernandez+Davis}).

In a reference frame $(x,\, y,\, z)$ with the $z$-axis along  $\boldsymbol{\hat{k}}$ and
such that the external magnetic field initially lies in the $x-z$
plane, the electric field associated to a photon of energy
$\hbar\omega$ can be written in terms of its complex amplitude
${\boldsymbol{A}}$ as
\begin{equation}
{\boldsymbol{E}}={\boldsymbol{E}}_0(z)e^{-i\omega
t}={\boldsymbol{A}}(z)e^{i(k_0z-\omega t)}\,,
\end{equation}
where $k_0=\omega/c$. In this frame ${\boldsymbol{A}}$ is in the
$x-z$ plane for a photon initially polarized in the O-mode, while
for an X-mode photon only the $y$ component of ${\boldsymbol{A}}$
is different from zero.
By solving the wave equation
\begin{equation} \label{eqn:waveequationgen}
{\boldsymbol{\nabla}}\times\left({\boldsymbol{\bar\mu}}\cdot{\boldsymbol{\nabla}}\times
{\boldsymbol{E}}\right)=\frac{\omega^2}{c^2}{\boldsymbol{\varepsilon}}\cdot{\boldsymbol{E}}\,,
\end{equation}
and retaining only linear terms, one obtains the following system
of differential equations for the complex amplitude which
determines the evolution of the polarization modes for radiation
propagating in a magnetized vacuum
\begin{equation} \label{eqn:AxAyequation}
\begin{aligned}
&\frac{\mathrm{d}A_{x}}{\mathrm{d}z} & = &\ \ \ \ \
\frac{ik_0\delta}{2}\left[MA_{x}+PA_{y}\right] \\
&\frac{\mathrm{d}A_{y}}{\mathrm{d}z} & = &\ \ \ \ \
\frac{ik_0\delta}{2}\left[PA_{x}+NA_{y}\right]\,,
\end{aligned}
\end{equation}
where $M$, $N$, $P$ and $\delta$ depend on the magnetic field (see the Appendix for the complete expressions)
\footnote{From the wave equation (\ref{eqn:waveequationgen}) it follows that $A_z=-(\varepsilon_{zx}A_x/\varepsilon_{zz}+
\varepsilon_{zy}A_y/\varepsilon_{zz})$. A non-vanishing $A_z$ implies that these are not plane waves. However, since it is
$\vert A_z\vert\ll \vert A_x\vert\sim\vert A_y\vert$, the amplitude of the oscillation along the propagation direction is
vanishingly small and will be neglected hereafter.}.

From equations (\ref{eqn:AxAyequation}) it is evident that the
scale length along which the complex amplitude varies is
$\ell_\mathrm{A}\sim 1/k_0\delta\propto B^{-2}$. This is to be compared with
the scale length $\ell_\mathrm{B}\sim
B/|{\boldsymbol{\hat{k}}}\cdot{\boldsymbol{\nabla}}B|$ along which the
external magnetic field varies. Near the star surface, where $B$
is higher, it is $\ell_{\mathrm{A}}\ll\ell_{\mathrm{B}}$. This means that the wave
electric field can instantaneously adapt its direction to that of
the magnetic field, which changes along the photon trajectory.
Under these conditions (adiabatic propagation), photons maintain
their initial polarization state, either O or X. As the photon
moves away from the star surface, $B$ decreases and $\ell_{\mathrm{A}}$
increases, until it becomes comparable to $\ell_{\mathrm{B}}$. The electric
field direction freezes and is not locked anymore to that of the local magnetic
field \cite[][]{hs00,hs02}. This occurs at a
characteristic distance, the polarization radius, which, for the
typical parameters of a magnetar, is $r_{\mathrm{pl}}\sim 150\, R_{\mathrm{NS}}$
(see \citeauthor{Fernandez+Davis}). Given that photons resonantly scatter up
to a radial distance $r_{\mathrm{esc}}\la 10\, R_{\mathrm{NS}}$ (see \S
\ref{subsec:magnetosphere}), it is $r_{\mathrm{pl}}\gg r_{\mathrm{esc}}$, which
makes it possible to treat the effects on polarization induced by
RCS and QED separately.

\section{Numerical simulations} \label{sec:theoreticalsimulations}

In order to compute the polarization properties of X-ray radiation
escaping from a magnetar magnetosphere, we follow closely the
approach described in \citeauthor{Fernandez+Davis}. RCS of primary
thermal photons is dealt with by means of the Monte Carlo code
developed by \citeauthor{NTZ1}, to which a new module was added to
solve the equations for the evolution of polarization modes in vacuo. The main
features of our numerical scheme, together with some illustrative runs,  are discussed in the following
subsections. Typical computing times are of about 30 minutes for processing $\sim 10^{6}$ photons on an Intel core i7 2.30 GHz processor.

\subsection{Monte Carlo code} \label{subsec:MonteCarlo}

Once the magnetospheric structure is fixed (polar value of the surface magnetic field $B_{\mathrm{p}}$,
twist angle $\Delta\phi_{\mathrm{N-S}}$, bulk velocity and temperature of the electrons, $\beta$
and $T_{\mathrm{el}}$), the code follows the propagation of photons, as they interact with the magnetospheric 
charges; general relativistic effects are not accounted for. Initially, photons are emitted from the 
cooling 
star surface with an assumed, isotropic blackbody distribution and arbitrary polarization state. The surface is divided into discrete, equal-area patches through an angular grid; each patch may have a different temperature. In the following, however, we take the temperature uniform (at $T$) on the whole surface\footnote{In the presence of an inhomogeneous temperature distribution, radiation from the hotter patches has tipically a larger polarization degree.} and assume that the seed photons are polarized in the X-mode, as discussed in \S \ref{subsec:polarization}. Since scatterings occur well inside the adiabatic zone, the photon polarization mode is held fixed between two successive scatterings, while it may change upon scattering (see again \S \ref{subsec:polarization}).

The code keeps track of the photon direction, energy and polarization state and when the escape condition is met
(i.e. the scattering probability becomes vanishingly small, see \citeauthor{ft07} 2007; \citeauthor{NTZ1}), integration of vacuum polarization
evolution is switched on. Actually, instead of equations (\ref{eqn:AxAyequation}), we found it more convenient on a
numerical ground to integrate the equations which govern the evolution of the Stokes parameters, $I$, $Q$, $U$ and
$V$. For monochromatic radiation, they are related to the components of the complex amplitude ${\boldsymbol{A}}$ in the reference frame introduced in \S \ref{subsec:polarization} by
\begin{equation}
\label{eqn:stokesdef}
\begin{aligned} &\bar I&=&\ \ \ \ \ A_{x}^{\ }A_{x}^*+A_{y}^{\ }A_{y}^* \\
&\bar Q&=&\ \ \ \ \ A_{x}^{\ }A_{x}^*-A_{y}^{\ }A_{y}^* \\
&\bar U&=&\ \ \ \ \ A_{x}^{\ }A_{y}^*+A_{y}^{\ }A_{x}^* \\
&\bar V&=&\ \ \ \ \ iA_{x}^{\ }A_{y}^*-iA_{y}^{\ }A_{x}^* \,,
\end{aligned}
\end{equation}
where a star denotes the complex conjugate. Moreover, they satisfy the general relation $\bar{I}^2\geq\bar{Q}^2+\bar{U}^2+\bar{V}^2$, where the
equality
holds for 100\% polarized radiation; the intensity $I$ is constant and, for a single photon, it can be taken as unity.
Hence, under our assumptions the initial conditions are simply given by $\bar U(0)=\bar V(0)=0$ and $\bar Q(0)=\pm 1$,
where the plus (minus) sign is for a photon initially in the X-mode (O-mode). However, when a large number of photons
is considered, as in our Monte Carlo simulations (see below), and the Stokes parameters are obtained by summation of
those relative to single photons, all the individual contributions must be referred to the same frame. Since we
collect photons propagating in the same direction (that is the line of sight), this amounts to select two fixed
directions normal to ${\boldsymbol{\hat{k}}}$.

\begin{figure*}
{\includegraphics[width=17.00cm]{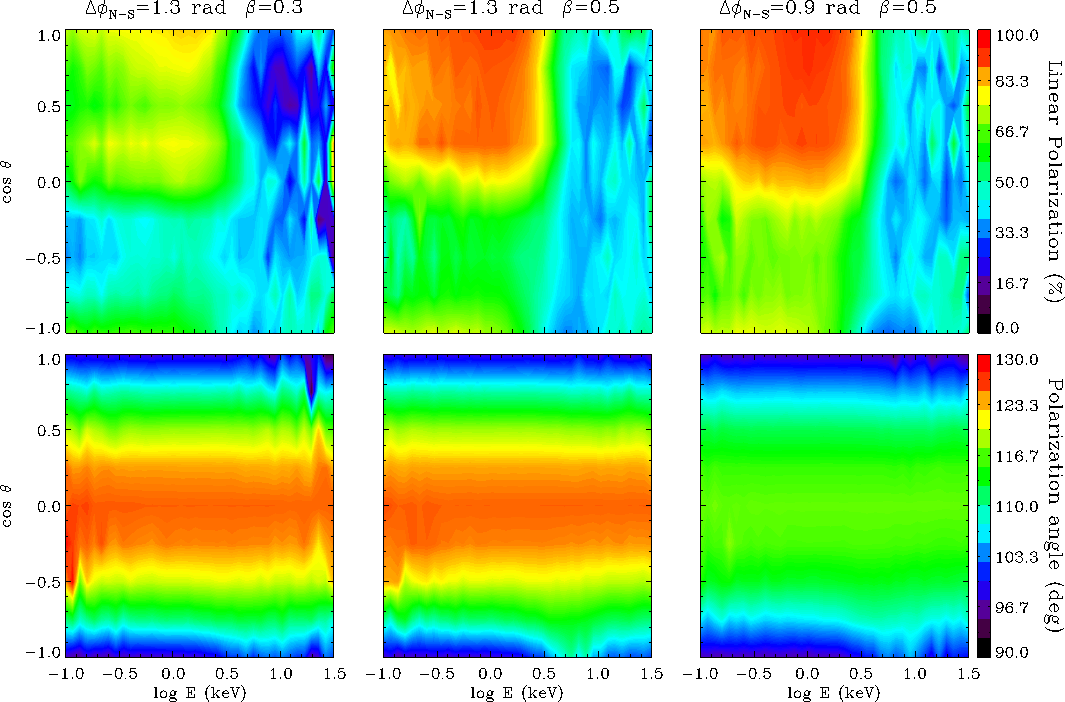}} \caption{\label{fig:PF+PAPhAv} Contour plots for the
polarization fraction (top row) and polarization angle (bottom row) as functions of the photon energy and $\cos\theta$
for different values of the twist angle and the electron bulk velocity: $\Delta\phi_\mathrm{N-S}=1.3$ rad, $\beta=0.3$
(left column); $\Delta\phi_\mathrm{N-S}=1.3$ rad, $\beta=0.5$ (middle column); $\Delta\phi_\mathrm{N-S}=0.9$ rad,
$\beta=0.5$ (right column). In all runs it is $B_\mathrm p=5\times 10^{14}$ G and $T_\mathrm{el}=10$ keV.}
\end{figure*}

In the new frame $(u,\, v,\, w)$, in which ${\boldsymbol{\hat{w}}}\equiv{\boldsymbol{\hat{k}}}$ and the
$u$-axis is perpendicular to both ${\boldsymbol{\hat{k}}}$ and the star spin axis $\boldsymbol{\hat{\Omega}}$, the
Stokes parameters are given by
\begin{equation} \label{eqn:stokesbaseu}
\begin{aligned}
&I&=&\ \ \ \ \ \bar I \\
&Q&=&\ \ \ \ \ \bar Q\cos(2\alpha)+\bar U\sin(2\alpha) \\ &U&=&\ \ \ \ \ \bar U\cos(2\alpha)-\bar Q\cos(2\alpha) \\
&V&=&\ \ \ \ \ \bar V \end{aligned} \end{equation} where
$\alpha=\arccos{\boldsymbol{\hat{u}}}\cdot{\boldsymbol{\hat{x}}}$ is the angle by which the new frame is rotated with
respect to the former around ${\boldsymbol{\hat{k}}}$.

It is easy to show that equations (\ref{eqn:AxAyequation}) are equivalent to \begin{equation} \label{eqn:stokesquation}
\begin{aligned} &\frac{\mathrm d Q}{\mathrm d z'}&=&\ \ \ \ \ -2PV \\ &\frac{\mathrm d U}{\mathrm d z'}&=&\ \ \ \ \
-(N-M)V\\ &\frac{\mathrm d V}{\mathrm d z'}&=&\ \ \ \ \ 2PQ+(N-M)U \,, \end{aligned} \end{equation} where $\mathrm d
z'=\kappa_0\delta\mathrm d z/2$. Some care must be taken in choosing the starting point for the integration of equations
(\ref{eqn:stokesquation}). Although the initial radius must be inside the adiabatic zone, including a large part of the
latter in the integration domain would be useless (the polarization mode does not change) and produce only an increase
in the computational time. After some experimenting (see also the discussion in \citeauthor{Fernandez+Davis}), we
decided to start the integration at a radial distance $r_{\mathrm{vac}}= 10^3\ell_{\mathrm{A}}$, where
$\ell_{\mathrm{A}}\simeq 100 (B/10^{11}\, \mathrm G)^{-2}(E/1 \, \mathrm{keV})^{-1}\, \mathrm{cm}$. Integration is
carried on until vacuum effects become negligible and the Stokes parameters freeze; this occurs at a radial distance $<
500R_{\mathrm{NS}}$, which we take as our fiducial upper bound.

Finally, escaping photons are collected on the sky at infinity,
i.e. on a spherical surface far enough that the NS appears
point-like. The sphere is divided into discrete patches by an
angular grid (much in the same way as the star surface when
dealing with thermal emission), each characterized by the magnetic
colatitude $\theta$ and azimuth $\phi$ of its centre. The program
returns, for each sky patch, the number of photons collected and
the Stokes parameters, sorted according to the energy; the latter
are computed by summing the values derived for the single photons.
The polarization observables are then computed as
\begin{equation} \label{eqn:PL+chipol}
\begin{aligned}
&\Pi_{\mathrm{L}} &=&\ \ \ \ \ \frac{\sqrt{Q^2+U^2}}{I}\,,
\\
&\chi_{\mathrm{pol}} &=&\ \ \ \ \
\frac{1}{2}\arctan\left(\frac{U}{Q}\right)\,,
\end{aligned}
\end{equation}
where $\Pi_{\mathrm{L}}$ is the linear
polarization fraction, i.e. the fraction of linearly polarized photons and $\chi_{\mathrm{pol}}$ is the polarization
angle, i.e. the angle between the $\boldsymbol{\hat{k}}-\boldsymbol{B}$ plane and the plane which contains the
oscillating electric field of the photons\footnote{The circular polarization fraction is not considered here because it
is not expected to be detectable with forthcoming X-ray instrumentation (see \S \ref{subsec:instrumentalsens}).}.

\subsection{Phase-averaged simulations} \label{subsec:phaseaveraged}

\begin{figure*} {\includegraphics[width=17.00cm]{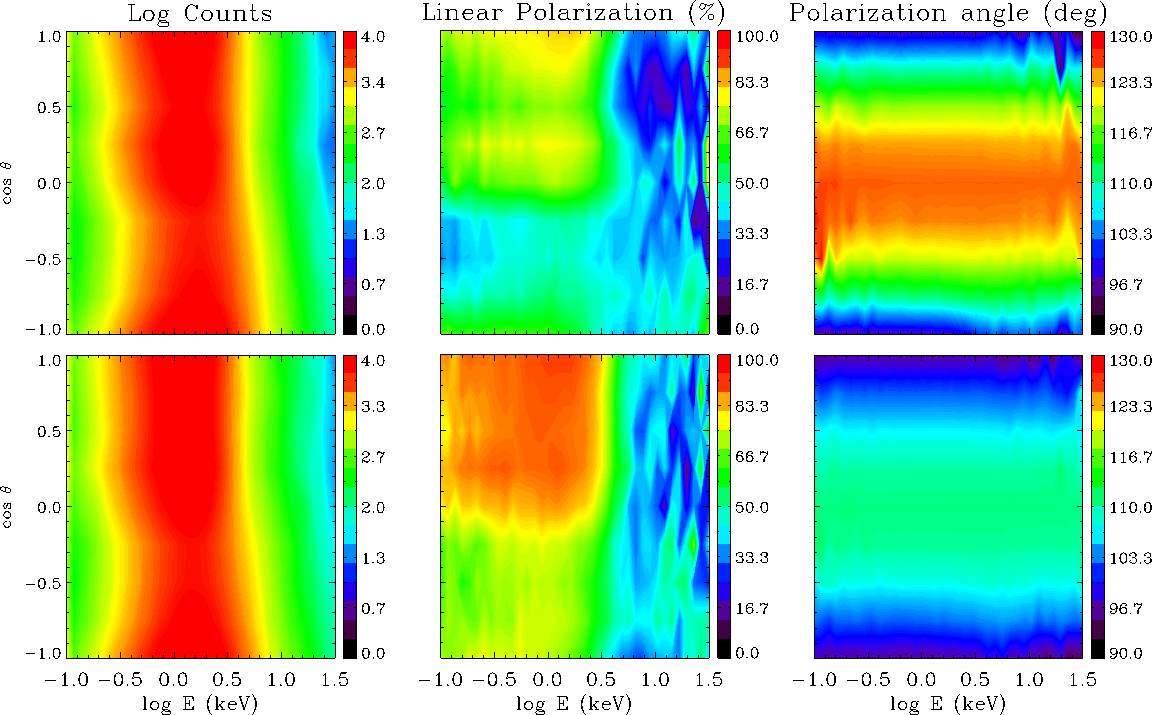}}
\caption{\label{fig:samespect} Contour plots for number of counts (in arbitrary units; left column), polarization fraction (middle column) and polarization angle (right column) as functions of the photon energy and $\cos\theta$ for different values of the twist angle and the electron bulk velocity: $\Delta\phi_\mathrm{N-S}=1.3$ rad, $\beta=0.3$ (top row) and $\Delta\phi_\mathrm{N-S}=0.7$ rad, $\beta=0.4$ (bottom row). In all runs it is  $B_\mathrm p=5\times 10^{14}$ G and $T_\mathrm{el}=10$ keV.}
\end{figure*}

To check our code and compare results with those obtained by \citeauthor{Fernandez+Davis}, we run first a number of
phase-averaged simulations. By introducing the direction of the line of sight (LOS; unit vector ${\boldsymbol{\hat
l}}$), the star viewing geometry is fixed by the two angles $\chi=\arccos{\boldsymbol{\hat
l}}\cdot{\boldsymbol{\hat{\Omega}}}$ and $\xi=\arccos{\boldsymbol{\hat{\Omega}}}\cdot {\boldsymbol{\hat{\mu}}}$, where
${\boldsymbol{\hat{\mu}}}$ and ${\boldsymbol{\hat{\Omega}}}$ are the unit vector along the magnetic and rotation axis,
respectively. For the sake of simplicity, in the following the star is taken to be an aligned rotator, i.e. $\xi=0$
so that $\chi=\theta$ (i.e. the LOS is fixed by the magnetic colatitude). Because of axial symmetry, data are averaged
with respect to the azimuthal angle $\phi$ in the reference frame of the star. All relevant quantities are then functions
only of the photon energy and of the magnetic colatitude.

Results for some typical runs are presented in Figure \ref{fig:PF+PAPhAv}, which shows the contour plots relative to
the polarization fraction $\Pi_\mathrm L$ (top row) and the polarization angle $\chi_\mathrm{pol}$ (bottom row) as
functions of energy and $\cos\theta$ for different values of the model parameters. In particular, by comparing the
left and middle columns the effects of changing the electron velocity, keeping all the other parameters fixed, can be
assessed. As it follows from equation (\ref{magcurr}), the electron density scales as $|\langle\beta\rangle|^{-1}$,
so for lower values of the electron bulk velocity their spatial density is higher, and photons undergo more
scatterings. As a result, the polarization degree is overall smaller (radiation is more depolarized) than for higher
$\beta$. On the other hand, the polarization angle does not change very much by varying the value of $\beta$. The
polarization fraction shows in addition a quite strong dependence on $\theta$ and $E$, which is evident in all the
three cases shown in Figure \ref{fig:PF+PAPhAv}. At low energies $\Pi_{\mathrm{L}}$ exhibits a clear asymmetry between
the northern and southern magnetic hemispheres (see also \citeauthor{Fernandez+Davis}). This behavior is due to the assumed
unidirectional flow of charged particles in the magnetosphere. Electrons stream from the north towards the south pole,
so that scatterings are more effective for photons coming from the the southern hemisphere (because collisions tend to
be more ``head on''), while those from regions above the magnetic equator retain more their initial polarization state
(here are 100\% polarized in the X-mode).

A comparison between the middle and right columns illustrates,
instead, the effects of varying the twist angle
$\Delta\phi_\mathrm{N-S}$, again with all other parameters held
fixed. Contrary to what happens by changing $\beta$, now the
variation affects both the polarization fraction and the
polarization angle. The effect on $\Pi_{\mathrm{L}}$ can be
understood by noticing that also the twist angle influences the
charge density (see equations \ref{eqn:Df} and \ref{magcurr}), so
when $\Delta\phi_{\mathrm{N-S}}$ is larger RCS is more efficient
and vice versa. On the other hand, the behavior of
$\chi_{\mathrm{pol}}$ appears to be quite independent on
scatterings: in all the three panels the polarization angle as a
function of energy is essentially constant, and deviations from
its initial value, $90^{\circ}$, are the same in the low and high
energy ranges. The polarization angle shows a stronger dependence
on the magnetosphere geometry (which is controlled by
$\Delta\phi_{\mathrm{N-S}}$), taking higher values as the twist
increases. More precisely (as already noticed by
\citeauthor{Fernandez+Davis}) it can be checked that
\begin{equation} \label{eqn:reltwistpol}
\chi_{\mathrm{pol}}=\arctan{\left(\frac{B_{\phi}}{B_{\theta}}\right)}+\frac{\pi}{2}\,.
\end{equation}

\begin{figure}
\begin{center}
{\includegraphics[width=8.00cm]{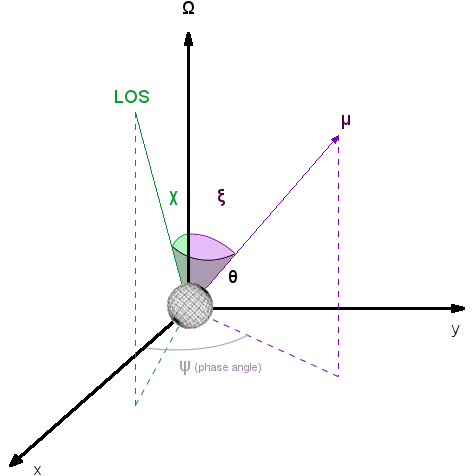}}
\caption{\label{fig:geomPR} Geometry for phase-resolved simulations.}
\end{center}
\end{figure}

\begin{figure*}
{\includegraphics[width=17.00cm]{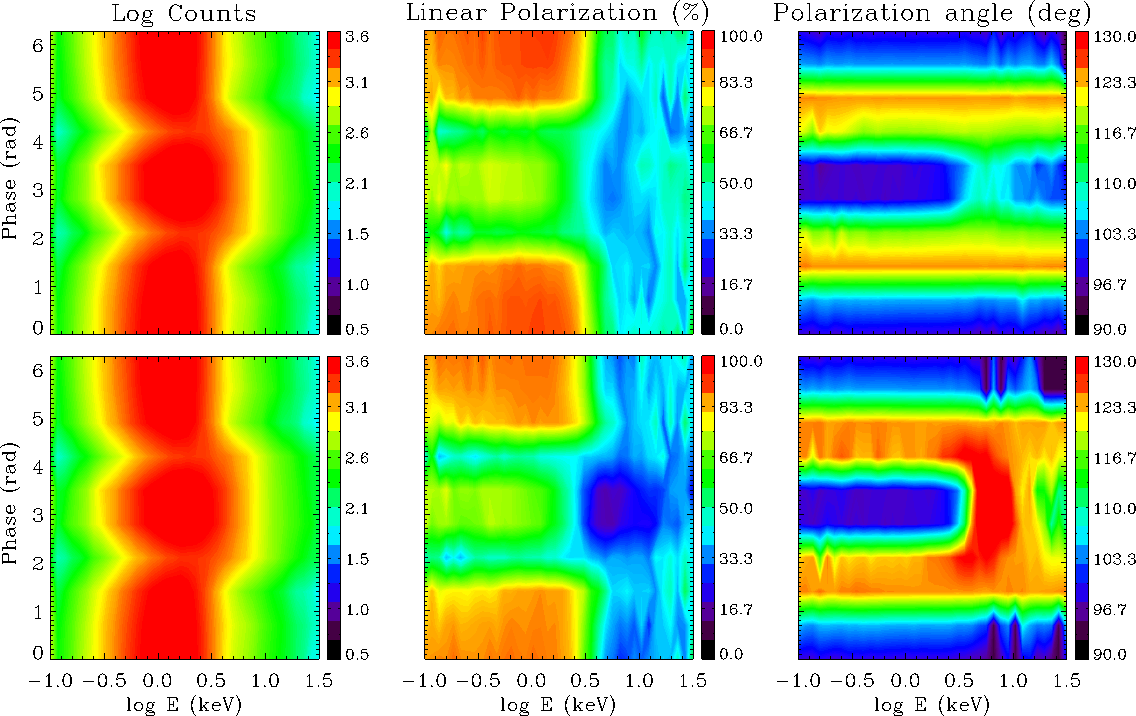}}
\caption{\label{fig:SP+PF+PAPhRs} Number of counts (in arbitrary units; left column), polarization 
fraction (middle column) and polarization angle (right column) as functions of energy and 
rotational phase, for a simulation with $B_p=4.6\times 10^{14}$ G, $\Delta\phi_\mathrm{N-S}=1.3$ rad, 
$\beta=0.5$ and $T_\mathrm{el}=10$ keV with (top row) and without (bottom row) QED effects. The star 
is assumed to be an orthogonal rotator seen perpendicularly to the spin axis ($\chi=\xi=90^{\circ}$).}
\end{figure*}

The only energy-dependent effect of scatterings on the
polarization angle is a small feature recognizable near the south
magnetic pole, between $\sim 3$ and 10 keV. This is also
associated to the north-south asymmetry we have already mentioned
(see again \citeauthor{Fernandez+Davis}). As a proof of the fact
that this feature is due to RCS, it tends to disappear for low
$\beta$ and becomes more evident for higher values. Finally we
checked that varying both $B_\mathrm{p}$ and $T_\mathrm{el}$ has a
very little effect on $\Pi_\mathrm L$ and $\chi_\mathrm{pol}$.

In order to illustrate the effectiveness of polarimetric
measurements in removing the degeneracy of the model, we performed
a series of simulations for different values of
$\Delta\phi_\mathrm{N-S}$ and $\beta$, in such a way to produce
spectra which are very close to each other. Results are shown in
Figure \ref{fig:samespect}, where the number of photons collected
at infinity (left column), polarization fraction (middle column)
and polarization angle (right column) are shown. Although the
plots for the photon spectrum are almost undistinguishable in the
two cases we report, $\Pi_\mathrm L$ and $\chi_\mathrm{pol}$ are
dramatically different. This actually proves that measurements of
polarization in magnetar X-ray emission could be of key importance
to probe the different geometries of the magnetosphere, in
addition to spectral analysis, which alone cannot, however,
suffice.

\subsection{Phase-resolved simulations} \label{subsec:phaseresolved}

In order to derive the variations of the polarization properties
with the star's rotational phase, we use the same method described
in \citeauthor{NTZ1} for computing the phase-resolved spectra and the
pulse profiles. As the star rotates, the angles $\theta$ and $\phi$ which the LOS makes with
the magnetic axis change in time according to (see Figure \ref{fig:geomPR})
\begin{equation} \label{eqn:thetachicsi}
\begin{aligned}
&\cos{\theta}&=&\ \ \ \ \ \cos\chi\cos\xi+\sin\chi\sin\xi\cos\psi \\
&\cos{\phi}&=&\ \ \ \ \ \frac{\cos{\chi}-\cos{\theta}\cos{\xi}}{\sin{\theta}\sin{\xi}}\,,
\end{aligned}
\end{equation}
where $\psi=2\pi t/P$ is the rotational phase ($P$ is the period).
This implies that
regions corresponding to different magnetic
colatitudes\footnote{Also the azimuth $\phi$ changes with the
phase, but this produces no effect because of the assumed symmetry
around the magnetic axis.} enter into view as the star rotates;
more precisely, the surface visibility range is
$\chi-\xi\leq\theta\leq\chi+\xi$ (see the first of equations
\ref{eqn:thetachicsi}). Once the angles $\chi$ and $\xi$ are
fixed, the position on the sky at infinity at which all the
(energy-dependent) quantities (the photon counts and the Stokes
parameters) are extracted is known for each value of the phase.  A
bilinear interpolation is actually used to obtain the values at
arbitrary positions on the sphere starting from those at the patch
centres.

An example of a typical phase-resolved output is shown in the top row of Figure \ref{fig:SP+PF+PAPhRs}, where the
photon spectrum, polarization fraction and polarization angle as functions of energy and rotational
phase are plotted for $\chi=\xi=90^{\circ}$, corresponding to an orthogonal rotator seen perpendicularly to the spin
axis. In this specific case, an observer can see all the surface, between the north (for $\psi=0,2\pi$) and the south
(for $\psi=\pi$) magnetic poles.

The polarization angle shows little dependence on the energy, as
expected from the phase-averaged results, again apart from the
feature localized near the south pole (i.e. around $\psi=\pi$ in
the present case) and between $\sim 3$ and $10$ keV, and due to
the assumed unidirectional flow, as already discusses in \S
\ref{subsec:phaseaveraged}. On the other hand, on varying the
phase, $\chi_{\mathrm{pol}}$ shows a maximum deviation from the
initial value of $90^{\circ}$ which occurs almost exactly at the
magnetic equator (seen twice, at $\psi=\pi/2$ and $3\pi/2$). Also
the behavior of the polarization fraction (top middle panel) is
rather similar to that of the phase-averaged simulations. However,
again for the asymmetry caused by the choice of the unidirectional
flow, in the $\sim 0.1$--3 keV range the minimum value of
$\Pi_{\mathrm{L}}$ occurs not in correspondence to the equator, as
for the maximum value of $\chi_{\mathrm{pol}}$, but just a bit
below (for $\psi\sim 2.1$ and $4.2$ rad). In fact, as noticed in
\S \ref{subsec:phaseaveraged}, $\Pi_{\mathrm{L}}$ is quite
sensitive to scattering, unlike $\chi_{\mathrm{pol}}$. This is the
main reason for which the behavior of the polarization angle is
more symmetrical between the northern and the southern hemispheres
than that of the polarization fraction, because the asymmetry is
related only to scatterings.


Phase-resolved simulations allow to clearly see the contribution of vacuum polarization effects (see \S
\ref{subsec:polarization}), as compared to those of RCS. The bottom row of Figure \ref{fig:SP+PF+PAPhRs} shows
again the  number of counts, polarization fraction and polarization angle for the same values of the model parameters
of the top row, but with the effects of QED turned off, so only RCS effects are accounted for. The photon spectrum is clearly the same, since it is not affected by vacuum polarization. The
plots concerning the polarization observables are, instead, substantially different, the most evident result being
that QED acts in smoothing out the polarization fraction and polarization angle behaviors. In particular, without
QED effects, $\chi_{\mathrm{pol}}$ (bottom right panel) shows a sharp dependence on energy near the south magnetic
pole where, as discussed above, RCS effects are more important. Also $\Pi_{\mathrm{L}}$ (bottom middle panel) is
affected by the absence of vacuum polarization, with an overall decrease of the polarization degree. Moreover,
the phase values at which the maximum of the polarization angle and the minimum of the
polarization fraction occur are closer to each other. So, in absence of vacuum polarization, the polarization
angle appears to be more sensitive to scatterings with respect to the complete QED+RCS
situation discussed above.

\section{Observability of the polarization signatures}

In the previous sections we have shown that polarimetry in X-rays can provide new observables for
studying the magnetosphere and constraining the geometrical angles in magnetars. Here, we are going to
investigate if, and to what extent, such observables can be measured by instruments which are likely  be
flown in the coming years on missions currently under development. To this end, we carried out detailed Monte Carlo
simulations for evaluating the response of an exemplary polarimeter to the
polarization signatures produced as radiation propagates through the magnetosphere. We first describe how
we calculate the sensitivity of the instrument and then we present how we derive, from a Monte Carlo
simulated measurement, the phase-resolved linear polarization degree and polarization angle.

\subsection{Instrumental sensitivity} \label{subsec:instrumentalsens}

Current polarimeters for X-ray astronomy are based on the
dependence of Bragg diffraction, photoelectric effect or Compton
scattering on the linear polarization of the incident radiation
since they can provide enough sensitivity for an astronomical
measurement. On the other hand, X-ray magnetic circular dichroism
and the dependence of Compton scattering on circular polarization,
have not been proven, as yet, of comparable efficiency; for this
reason, and given the low degree of circular polarization expected
in magnetars ($\la 5\%$), in the following circular polarization
will not be considered. We focus our discussion on the 2--6 keV
energy range because the spectrum of magnetar sources peaks around
a few keV's, here the polarization signatures are more evident and
the measurement is easier to accomplish. Moreover, since AXPs/SGRs
are relatively faint sources at least in quiescence, the use of an
X-ray telescope is usually convenient and, in this energy range,
conventional telescopes based on grazing incidence can be easily
exploited.

In this X-ray range the most promising polarimeters are those
based on the photoelectric effect \citep{Costa2001}. They can
measure the polarization of the beam together with its spectrum
with a moderate energy resolution, of the order of 20\% at 6~keV,
and with an accurate timing of the event, usually at the level of
few microseconds \citep{Bellazzini2006,Black2007}. In addition to
that, the Gas Pixel Detector
\citep[GPD,][]{Bellazzini2007,Bellazzini2010c} can provide also
very good imaging capabilities \citep{Soffitta2013b, Fabiani2013},
which are particularly useful for studying faint sources because
this allows for a proper removal of the background. Therefore, we
will discuss in the following the sensitivity of an instrument
based on the GPD, which is however quite representative of this
class of instruments.

The GPD has been presented as focal plane detector in a number of mission proposals, together with
small \citep{Costa2010c}, medium \citep{Tagliaferri2012} or large \citep{Bellazzini2010b} area
telescopes. In the following, we will take as an example the small mission XIPE
\citep{Soffitta2013}, recently proposed to the European Space Agency in the context of a call for
launch in 2017, to prove that even a mission with limited resources can be extremely useful in studying the
magnetospheric environment of a magnetar. We use a Monte Carlo technique to derive
the value of the polarization which would be measured by the instrument and its error. The code
has been already described in detail \citep{Dovciak2011} and here we summarize only its
most relevant features.

In general, polarization in X-rays is derived from the measured
\emph{modulation curve}, which is basically the histogram of the
azimuthal response of the instrument. For example, the modulation
curve for photoelectric polarimeters is the histogram of the
azimuthal emission direction of the photoelectrons
\citep{Bellazzini2010}. In case of polarized photons, the
modulation curve shows a cosine square modulation the phase of
which is related to the polarization angle and coincides with it
for photoelectric polarimeters. The amplitude of the modulation is
proportional to the degree of polarization and to the
\emph{modulation factor} $\mu$, which is the amplitude of the
instrumental response to completely polarized photons. 

The purpose
of the Monte Carlo is to produce a number of ``trial'' modulation
curves in the energy range of interest, fit them with a cosine
square function and derive for each trial an estimate of the
polarization which would be measured from that modulation curve. The number of entries in the histogram is instead the number of
collected events in the considered energy interval, obtained by
multiplying the source spectrum by the collecting area of the
telescope and by the instrument efficiency, using the response
matrix of the instrument including its energy resolution. Each
trial is affected by a different Poisson noise in the number of
entries per azimuthal beam; systematic effects, proven to be lower
than 1\% for the GPD \citep{Bellazzini2010c}, are neglected. In
defining the energy interval, the code takes into account the
finite energy resolution of the instrument. The ``measured'' angle
and degree of polarization which are provided by the Monte Carlo
are the values derived by a random trial, whereas their errors are
the average values over all trials. The efficiency and the
modulation factor of the GPD are discussed in detail in other
papers to which the interested reader is referred for more
information \citep{Muleri2008,Muleri2010}, whereas the collecting
area of XIPE is presented in \citet{Soffitta2013}.

\subsection{Simulated polarization measurements} \label{subsec:simpolmeas}

As discussed in Sec. \ref{sec:theoreticalsimulations}, the
polarization signature of magnetars depends on a number of parameters. Here we
aim at investigating if X-ray polarimetry can be exploited to
measure them and to which extent observations can discriminate
between different cases. In the following we make explicit
reference to phase-resolved measurements, which  are the most
promising because, albeit a phase-averaged measurement integrates
all the counts, its expected degree of polarization is smaller
since the polarization angle swings across the rotational phases.

The sensitivity of the XIPE mission is evaluated using as a template the AXP
1RXS J170849.0-400910 (\src\ for short)\footnote{See the McGill
on-line magnetar catalogue at
http://www.physics.mcgill.ca/~pulsar/magnetar/main.html and references therein.}.
The source period and period derivative are $P\simeq 11\, \rm
s$ and $\dot P\simeq 1.9\times 10^{-11}\, {\rm ss}^{-1}$, respectively, implying a dipole field of $4.6\times 10^{14}\,
\rm G$. \src, one of the brightest known magnetars \citep{Rea2005,Campana2007}, is slightly variable, with 
a (unabsorbed) flux ranging between 21 and $35\times10^{-12}\, \flux$ when restricted to the 
2--6 keV band. The estimated source distance is $\sim 3.8$~kpc \citep{dkvk06}
and we adopt the column density derived by \citet{Rea2005},
$N_\mathrm{H}=1.48\times10^{22}\, {\rm cm}^{-2}$. Spectral fits to
high-statistics \emph{XMM-Newton} data of \src\ with the XSPEC NTZ
model (i.e. the same spectral model discussed in Section
\ref{sec:theoreticalmodel}) have been presented in \cite{zrtn09}.
The best fit parameters are: $kT=0.47\, {\rm keV}$, $\beta=0.34$,
$\Delta\phi_\mathrm{N-S}=0.49$; the column density,
$N_\mathrm{H}=1.45\times10^{22}\, {\rm cm}^{-2}$, is fully in
agreement with that by \cite{Rea2005}. No estimate of the angles
$\xi$ and $\chi$ could be derived, since the NTZ model is
angle-averaged, nor it  has been obtained by other means.

The study of magnetar sources with similar properties to \src\ would be in the core science for a small mission
dedicated to X-ray polarimetry like XIPE. Thus we assume a total observation time of 1~Ms, which is completely
reasonable for such a kind of mission. The simulated source photon spectrum is then generated, as discussed in Section
\ref{subsec:instrumentalsens}, starting from the output of the Monte Carlo code (see Section \ref{subsec:MonteCarlo})
for a given set of parameters. In order to produce phase-resolved polarization observables, data are collected in nine,
equally-spaced phase bins. In the following we
present simulations obtained for a model with the same magnetospheric parameters as derived from the spectral fit of
\src\ (see above), together with a set of other test cases, obtained varying $\Delta\phi_\mathrm{N-S}$ and $\beta$. In
all simulations the magnetic field and the column density were held fixed at the values inferred for \src\ and model
spectra have been
normalized in such a way to produce a flux comparable to that of \src\ for the assumed distance of 3.8 kpc. The electron
temperature is always set at $T_\mathrm{el}= 10\, \mathrm{keV}$.

\begin{figure}
\begin{center}
\includegraphics[width=8cm]{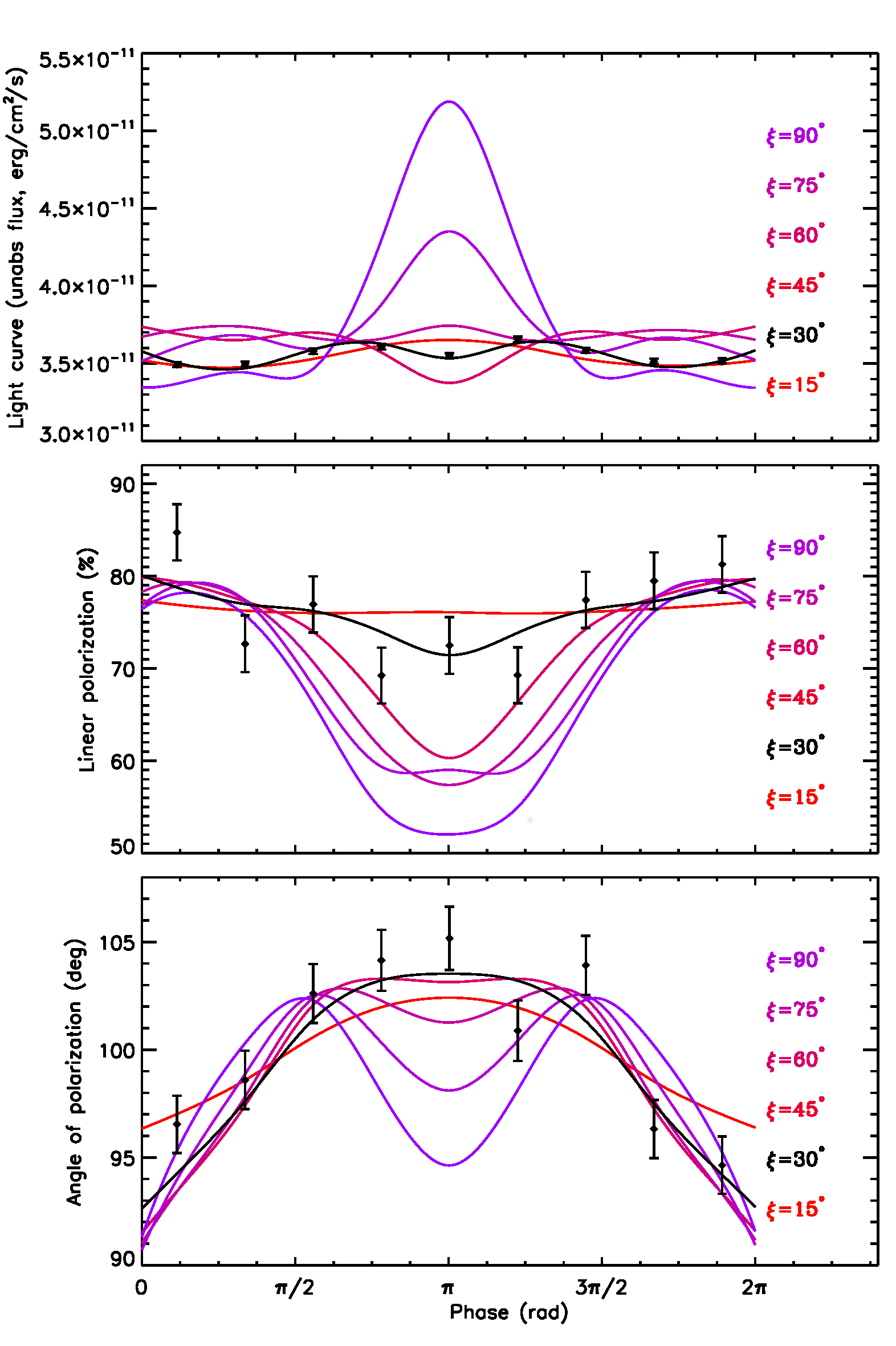}
\caption{Pulse profile (top panel),  phase-resolved polarization
degree (middle panel) and polarization angle (bottom panel) for a
set of models with the same viewing direction ($\chi=60^\circ$)
but with a different values of $\xi$. The filled circles are the simulated data (with the associated
1 $\sigma$ errors) obtained with a 1~Ms observation
of \src\  by XIPE assuming $\chi=60^\circ$ and $\xi=30^\circ$. All quantities refer to the 2--6 keV 
energy range.}
\label{fig:Geometry}
\end{center}
\end{figure}

A first example is shown in Figure \ref{fig:Geometry}, where the
XIPE simulated data for $kT=0.47\, {\rm keV}$, $\beta=0.34$,
$\Delta\phi_\mathrm{N-S}=0.49$, $\chi=60^\circ$, $\xi=30^\circ$
are compared with the model. The three panels refer to the 2--6
keV pulse profile (top), linear polarization fraction (middle) and
polarization angle (bottom). The filled circles with error bars
are the XIPE measured quantities while the solid lines represent
the models computed for the same parameter values and different
values of $\xi$ (the model from which the simulated data were
derived is the black curve). As a matter of fact, since the
polarization angle is almost independent on the energy, we can
derive the average degree of polarization of each model in the 2--6
keV energy range by simply weighting the phase-resolved polarization spectrum
$P(E,\psi)$ with the photon spectrum $S(E,\psi)$, that is,
$P(\psi)_{\mathrm{2-6\, keV}} = \int_\mathrm{2\, keV}^\mathrm{6\, keV} P(E,\psi)
S(E,\psi)\, dE/\int_\mathrm{2\, keV}^\mathrm{6\, keV} S(E,\psi)\, dE$. Although this issue will be
discussed in more detail later on, it is quite evident that a
\emph{simultaneous} fit of the degree and angle of polarization
would allow to derive unambiguously the value of $\xi$; additional
information could be derived from the light curve that the
instrument provides thanks to its very good timing properties. A
similar result holds by keeping constant the angle $\xi$ and
letting $\chi$ free to vary.

\begin{figure}
\begin{center}
\includegraphics[width=8cm]{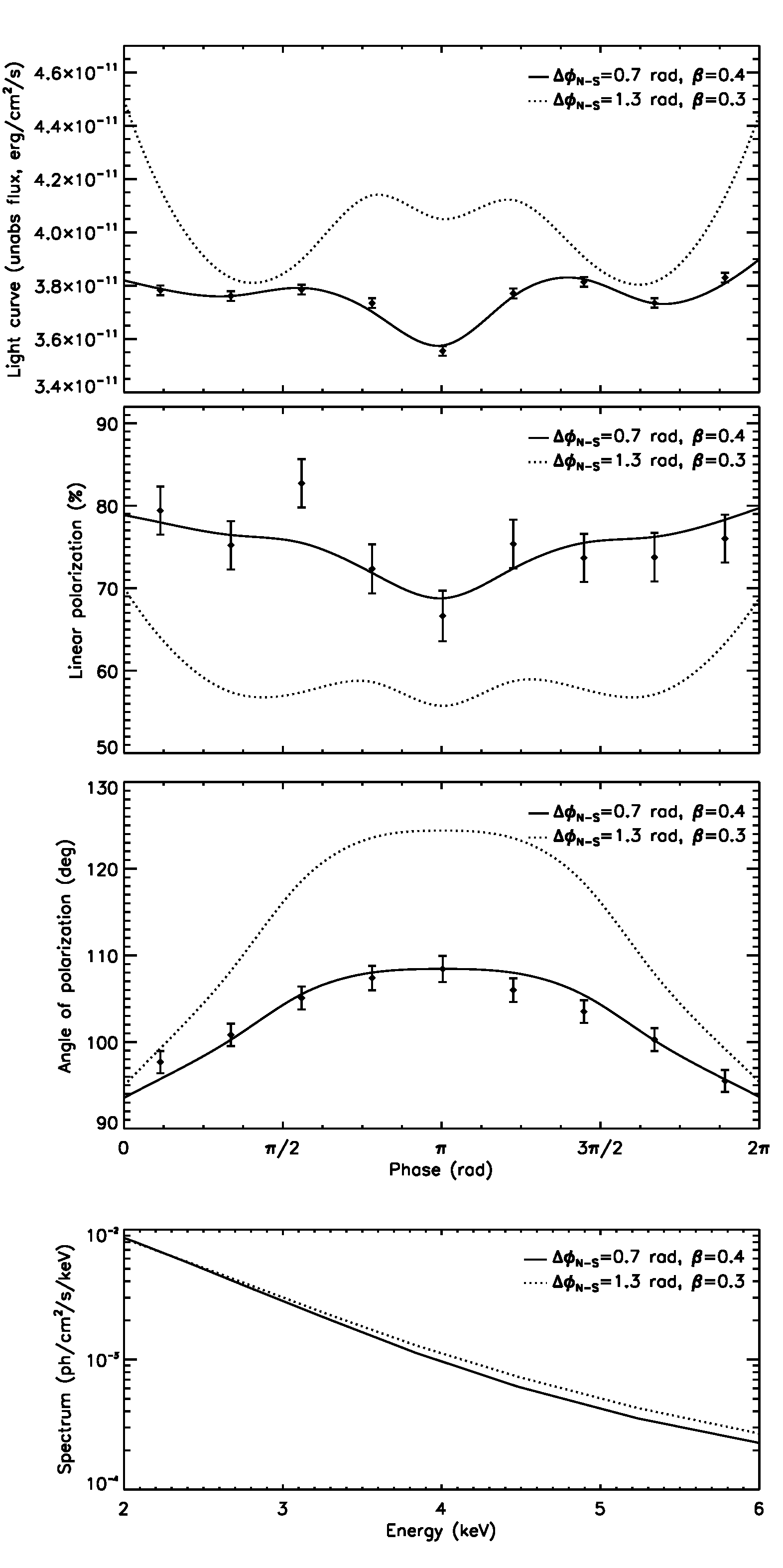}
\caption{From top to bottom: comparison of the pulse profile,
polarization degree, polarization angle and photon spectrum for two
models characterized by $\Delta\phi_\mathrm{N-S}=0.7\mathrm{\ rad},\,
\beta=0.4$ (solid line, model A) and
$\Delta\phi_\mathrm{N-S}=1.3\mathrm{\ rad},\, \beta=0.3$ (dotted line, model
B). The filled circles with error bars denote the simulated XIPE
data obtained from model A. All quantities refer to the 2--6 keV energy range.} \label{fig:Parameters}
\end{center}
\end{figure}

As a second example, which well illustrates the merits of X-ray
polarimetry, we consider a case in which a different set of
parameters produce an almost indistinguishable spectrum (this is
the degeneracy which we already mentioned). The two models are
shown in Figure~\ref{fig:Parameters}: one has
$\Delta\phi_\mathrm{N-S}=0.7$~rad, $\beta=0.4$ (solid line), the
other $\Delta\phi_\mathrm{N-S}=1.3$~rad and $\beta=0.3$ (dotted
line); both models are for $\xi=60^\circ$ and $\chi=30^\circ$. The
simulated XIPE data (filled circles with error bars) were
generated from the former. As it can be seen from the bottom
panel, the unabsorbed spectra in the 2--6 keV range are almost
identical, whereas a 1~Ms observation with XIPE clearly sets the
two cases apart, both as far as the polarization degree and angle
are concerned. It is worth noting that also a precise fitting of
pulse profile allows to discriminate the two models and in fact
this approach has been already put into practice \citep[see, e.g.,
][]{alba10}. It is beyond the scope of this paper to determine
whether polarimetry or phase-resolved photometry are more
sensitive in measuring the magnetospheric parameters. Here we just
say that the addition of these two new observables can be crucial
for assessing the consistency of the model and, if so, to enhance
the significance of the measurement.

\begin{table*}
\centering
\caption{Best fit parameters$^{\rm a}$}
\label{table:fitparam}
\begin{tabular}{@{}lccccc}
\hline
 & $\beta$ & $\Delta\phi_\mathrm{N-S}$ & $\chi$ & $\xi$ & $\chi^2_\mathrm{red}$\\
 & & rad & deg & deg & \\
\hline
Input values & $0.34$ & $0.5$ & $60$ & $30$ & --\\
\src & $0.34\pm 0.004$ & $0.51\pm 0.01$ & $61.4\pm 0.9$ & $25.7\pm 1.8$ & 0.97\\
\hline
Input values & $0.5$ & $1.3$ & $90$ & $60$ & --\\
QED-on & $0.501\pm 0.004$ & $1.17\pm 0.03$ & $90.3\pm 1.3$ & $57.9\pm 2.2$ & 1.17\\
QED-off & $0.503\pm 0.002$ & $1.34\pm 0.04$ & $89.3\pm 1.4$ & $60.4\pm 3.2$ & 7.78\\
\hline
\end{tabular}
\begin{list}{}{}
\item[$^{a}$]Reported errors are at $1\sigma$ level.
\end{list}
\end{table*}

\begin{figure*}
\begin{center}
\includegraphics[width=17cm]{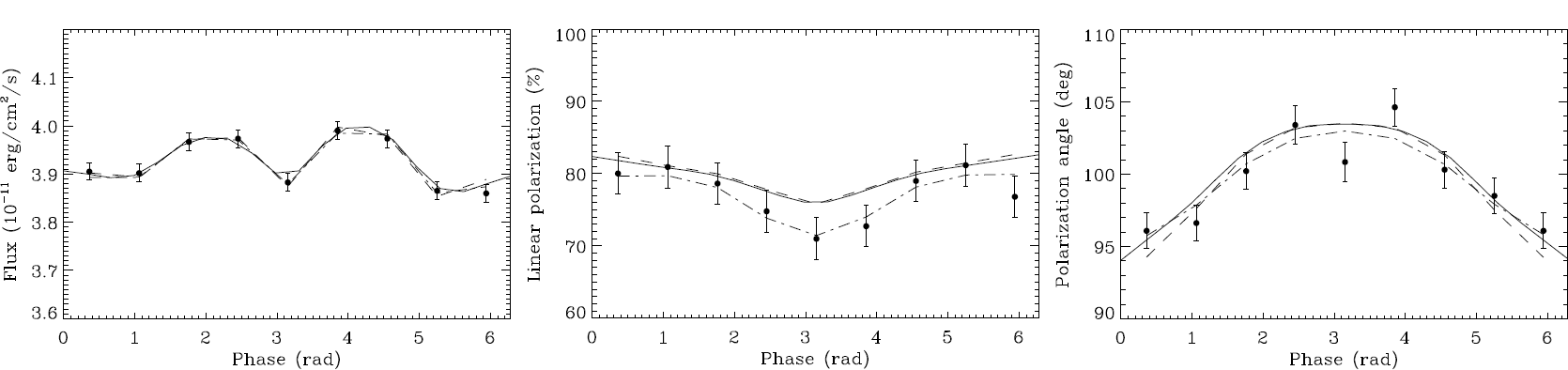}
\caption{Simulated data (filled circles) and best simultaneous fit of the pulse profile, polarization degree and
polarization angle in the $2$--6 keV energy range (full lines) for the same model as in figure \ref{fig:Geometry}
($\beta=0.34$, $\Delta\phi_\mathrm{N-S}=0.5\, \mathrm{rad}$,
$\chi=60^\circ$ and $\xi=30^\circ$). The dashed and dash-dotted lines show, respectively,
the model from which data were generated and the individual fits.}
\label{fig:1708}
\end{center}
\end{figure*}

Finally, we discuss the capability of the method to recover the
input model parameters from fitting the simulated data without any
assumption, much in the same way as it would be done when
confronting the model with ``real'' observational data. To this
end we exploit all the available observables, pulse profile,
phase-dependent linear polarization fraction and angle, and
perform a simultaneous fit leaving both the magnetospheric
parameters ($\beta$, $\Delta\phi_\mathrm{N-S}$) and the
geometrical angles ($\chi$, $\xi$) free to vary. The magnetic
field strength, the surface and electron temperature together with
the column density are, instead, held fixed at the values
introduced previously and we refer here to the same case
illustrated in figure \ref{fig:Geometry}, which is representative
of \src. In this respect, we note that $B$ can be estimated from the
spin-down measure and both $kT$ and $N_\mathrm H$ from spectral
fitting. Much in the same way as for the XSPEC NTZ model \cite[see
NTZ, ][for details]{zrtn09}, we produced beforehand a model
archive covering the parameter ranges $0.2\leq\beta\leq 0.7$ (step 0.1), $0.3\, \mathrm{rad}\leq\Delta\phi\leq
1.4\, \mathrm{rad}$ (step 0.1), $15^\circ\leq\chi\leq 150^\circ$ (step $15^\circ$) and $15^\circ\leq\xi\leq
90^\circ$ (step $15^\circ$).
Single models were then
loaded in an IDL script which performs the fitting exploiting
linear interpolation to obtain models for values of the parameters
not included in the archive. We performed both simultaneous fits
(pulse profile + polarization fraction + polarization angle) and
individual fits to single observables. Results are shown in Figure
\ref{fig:1708} and summarized in Table \ref{table:fitparam}. The simultaneous fit of all the three observables
is quite satisfactory ($\chi^2_\mathrm{red}=0.97$) and returns parameter values which are well compatible with
the input ones at the $2\sigma$ level. The only exception of the angle $\xi$, for which a value somewhat lower
than the input one is derived (see again Table \ref{table:fitparam}).

\begin{figure*}
\begin{center}
\includegraphics[width=17cm]{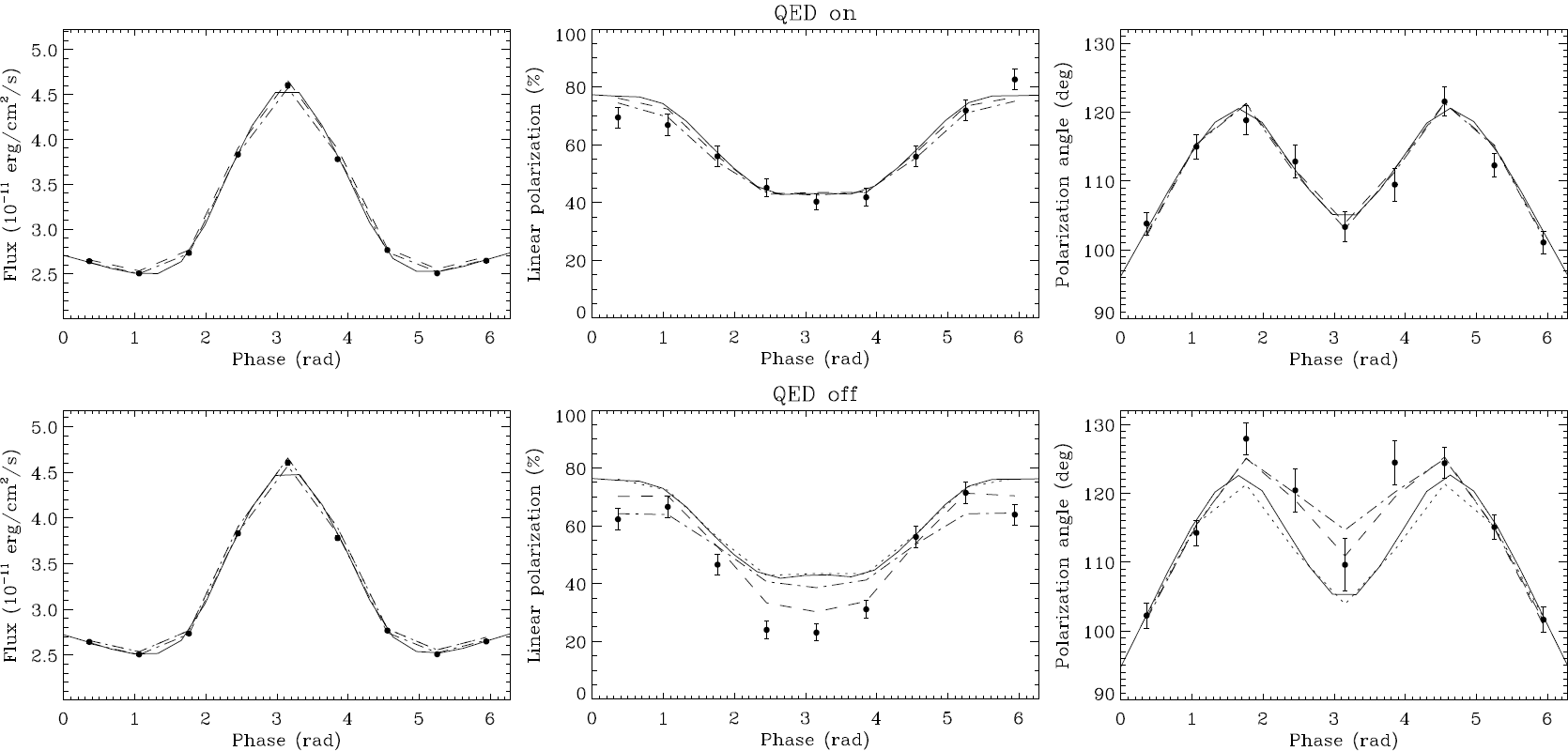}
\caption{Comparison of the pulse profile, polarization degree and
polarization angle (again in the $2$--6 keV energy range) for two models both characterized by
$\Delta\phi_\mathrm{N-S}=1.3\, \mathrm{rad}$, $\beta=0.5$,
$\chi=90^\circ$ and $\xi=60^\circ$. The
filled circles with error bars are the XIPE data, generated with (top row) and without (bottom row) vacuum polarization. The full and dash-dotted lines are the simultaneous and individual fits, respectively, with ``QED-on'' models.
The dashed lines show the models from which data were produced, with (upper panels) and without (bottom panels) QED effects; in the bottom panels the model with
vacuum polarization is also shown for comparison (dotted lines).}
\label{fig:QEDonoff}
\end{center}
\end{figure*}

We have seen that, at variance with the spectrum, the polarization
signature is quite dependent on vacuum polarization. To better
illustrate this, we computed two models for the same values of the
parameters ($\chi=90^\circ$, $\xi=60^\circ$,
$\Delta\phi_\mathrm{N-S}=1.3$~rad and $\beta=0.5$, similar to
those of figure \ref{fig:SP+PF+PAPhRs}), the first with both
scattering and vacuum polarization taken into account
(``QED-on''), whereas in the second vacuum polarization was turned
off (``QED-off''). We generated XIPE data from both simulations
and attempted to fit the two sets using the model archive we
introduced above and which includes vacuum polarization. Results
are shown in Figure~\ref{fig:QEDonoff}, where the top panels refer
to the ``QED-on'' case and the bottom panels to the ``QED-off''
one. While, as expected, the pulse profiles are indistinguishable,
the polarization observables are quite different in the two cases,
as it can be seen from both the data points and the generating
models (dashed lines). For ``QED-on'' the simultaneous fit of the
pulse profile, polarization fraction and angle is satisfactory
($\chi^2_\mathrm{red}\sim 1.2)$ and returns values compatible,
within the errors, with the input ones (see Table \ref{table:fitparam}). This,
however, does not occur for the  ``QED-off'' data, to which
``QED-on'' models provide an unacceptable representation
($\chi^2_\mathrm{red}\sim 7.8)$. This shows that polarimetric
measures are potentially capable of pinpointing QED effects in
magnetar magnetospheres and can provide an indirect evidence of an
ultra-strong magnetic field.

\section{Conclusions}

Previous investigations (Fern\'{a}ndez \& Davis 2011; see also Nobili, Turolla \& Zane 2008)
have clearly indicated the potentialities  of polarimetric
measures at X-ray energies in probing the physics of magnetars. In
this investigation we have reconsidered the issue of polarization
of X-ray radiation from magnetars within the framework of the
twisted magnetosphere model, which has been successfully applied
to reproduce the soft X-ray spectra of several magnetar candidates
(SGRs/AXPs). While retaining some simplifications (globally
twisted magnetosphere and unidirectional flow of charged particles
along the closed field lines), our Monte Carlo simulations
properly account for both the effects of resonant cyclotron
scattering and of QED on the final polarization state of the
emergent radiation.

Phase-averaged as well as phase-resolved results confirm that the
linear polarization fraction $\Pi_{\mathrm{L}}$ and the
polarization angle $\chi_{\mathrm{pol}}$, are very sensitive to
the magnetospheric twist angle $\Delta\phi_{\mathrm{N-S}}$ and the
charge velocity $\beta$, and also to the geometric angles $\chi$
and $\xi$. This allows to remove the
$\Delta\phi_{\mathrm{N-S}}$-$\beta$ degeneracy which spectral
measures alone can not disambiguate. Polarization measurements can
also univocally discriminate between cases in which QED effects
are present or not.

In order to assess the feasibility of magnetar X-ray polarization measurements, we simulated a 1 Ms 
observation of the bright AXP 1RXS J170849.0-400910 with XIPE, a small X-ray polarimetry mission 
recently proposed. 
We showed that it is indeed  possible to extract the values of the physical and geometrical parameters from a phase-resolved 
measurement of polarization observables, by fitting the simulated data with a large set of our theoretical models. This more in-depth analysis allows also to distinguish between different configurations in which photon spectra are 
very similar, therefore removing possible degeneracies. 
Finally, we proved that polarimetric measures are sensitive enough to reveal QED effects due to vacuum polarization, 
showing that polarization can be used as a tool to confirm 
the presence of ultra-strong magnetic fields in magnetars.
Indeed, the sensitivity of state-of-the-art X-ray polarimeters appears today perfectly adequate to
successfully probe the magnetospheric environment of a magnetar, ideally complementing spectral and timing measures and
providing invaluable insight into the physical processes at the basis of ultra-magnetized neutron stars.

\section*{Acknowledgments}
We thank Silvia Zane and Enrico Costa for a number of useful comments on the
manuscript. RT acknowledges financial support from an INAF PRIN 2011 grant.

\appendix{} \label{sec:appendix}
\section{The coefficients $M$, $N$ and $P$} \label{subsec:appMNP}
Since the polarization radius is in a region where $B\ll B_{\mathrm{Q}}$ 
($B_{\mathrm{Q}}=m_e^2c^3/(\hbar e)\simeq 4.414\times 10^{13}$ G is the quantum critical field), 
the complete expressions of the coefficients $M$, $N$ and $P$ which appear in equations (\ref{eqn:AxAyequation})
and (\ref{eqn:stokesquation}) are,  in the weak-field limit,
\begin{equation} \label{eqn:completeMNP}
\begin{aligned}
&M&=&\ \ \ \ \ \frac{(7\hat{B}^2_x+4\hat{B}^2_y)\bar{\mu}_{xx}-12\delta\hat{B}^2_x\hat{B}^2_y}{\bar{\mu}_{xx}\bar{\mu}_{yy}-16\delta^2\hat{B}^2_{x}\hat{B}^2_{y}} \\
&N&=&\ \ \ \ \ \frac{(4\hat{B}^2_x+7\hat{B}^2_y)\bar{\mu}_{yy}-12\delta\hat{B}^2_x\hat{B}^2_y}{\bar{\mu}_{xx}\bar{\mu}_{yy}-16\delta^2\hat{B}^2_{x}\hat{B}^2_{y}} \\
&P&=&\ \ \ \ \ \frac{[3\bar{\mu}_{xx}-4\delta(7\hat{B}^2_y+4\hat{B}^2_x)]\hat{B}_x\hat{B}_y}{\bar{\mu}_{xx}\bar{\mu}_{yy}-16\delta^2\hat{B}^2_{x}\hat{B}^2_{y}}\,.
\end{aligned}
\end{equation}
The components of the inverse permeability tensor $\boldsymbol{\bar{\mu}}$ are given in equation
(\ref{eqn:epsilonmu}) and
\begin{equation} \label{eqn:completedelta}
\begin{aligned}
&\delta &=&\ \ \ \ \ \frac{\alpha_{\mathrm{F}}}{45\pi}\left(\frac{B}{B_{\mathrm{Q}}}\right)^2 \\
&\ &\simeq &\ \ \ \ \  3\times 10^{-10}\left(\frac{B}{10^{11}\mathrm{\ G}}\right)^2\,,
\end{aligned}
\end{equation}
where $\alpha_{\mathrm{F}}\simeq 1/137$ is the fine-structure constant (see \citeauthor{Fernandez+Davis}).

\label{lastpage}


\addcontentsline{toc}{chapter}{Bibliografia}
\begin{thebibliography}{}
\bibitem[\protect\citeauthoryear{Albano et al.}{2010}]{alba10}
Albano A., Turolla R., Israel G.L., Zane S., Nobili L., Stella L. 2010, ApJ, 722, 788

\bibitem[{{Bellazzini} {et~al}\mbox{.}(2010){Bellazzini}, {Brez}, {Minuti},
  {Pinchera}, {Spandre}, {Muleri}, {Costa}, {Di Cosimo}, {Fabiani},
  {Lazzarotto}, {Rubini}, \& {Soffitta}}]{Bellazzini2010b}
{Bellazzini} R. {et~al.} 2010, in X-ray Polarimetry: A New Window in
  Astrophysics, Cambridge University Press

\bibitem[{{Bellazzini} \& {Muleri}(2010)}]{Bellazzini2010c}
{Bellazzini} R., {Muleri} F. 2010, Nuclear Instruments and Methods in Physics
  Research A, 623, 766

\bibitem[{{Bellazzini} \& {Spandre}(2010)}]{Bellazzini2010}
{Bellazzini} R., {Spandre} G. 2010, in X-ray Polarimetry: A New Window in
  Astrophysics, Cambridge University Press

\bibitem[{{Bellazzini} {et~al}\mbox{.}(2007){Bellazzini}, {Spandre}, {Minuti},
  {Baldini}, {Brez}, {Latronico}, {Omodei}, {Razzano}, {Massai},
  {Pesce-Rollins}, {Sgr{\'o}}, {Costa}, {Soffitta}, {Sipila}, \&
  {Lempinen}}]{Bellazzini2007}
{Bellazzini} R. {et~al.} 2007, Nuclear Instruments and Methods in Physics
  Research A, 579, 853

\bibitem[{{Bellazzini} {et~al}\mbox{.}(2006){Bellazzini}, {Spandre}, {Minuti},
  {Baldini}, {Brez}, {Cavalca}, {Latronico}, {Omodei}, {Massai},
  {Sgr{\'o}}, {Costa}, {Soffitta}, {Krummenacher}, \&
  {de Oliveira}}]{Bellazzini2006}
{Bellazzini} R. {et~al.} 2006, Nuclear Instruments and Methods in Physics
  Research A, 566, 552

\bibitem[\protect\citeauthoryear{Beloborodov \& Thompson}{2007}]{bt07}
Beloborodov A.M., Thompson C. 2007, ApJ, 657, 967

\bibitem[\protect\citeauthoryear{Beloborodov}{2009}]{belo09}
Beloborodov A.M. 2009, ApJ, 703, 1044

\bibitem[\protect\citeauthoryear{Beloborodov}{2013}]{belo13}
Beloborodov A.M. 2013, ApJ, 764, 157

\bibitem[{{Black} {et~al}\mbox{.}(2007){Black}, {Baker}, {Deines-Jones},
  {Hill}, \& {Jahoda}}]{Black2007}
{Black} J.~K., {Baker} R.~G., {Deines-Jones} P., {Hill} J.~E., {Jahoda} K.
  2007, Nuclear Instruments and Methods in Physics Research A, 581, 755

\bibitem[{{Campana} {et~al}\mbox{.}(2007){Campana}, {Rea}, {Israel}, {Turolla},
  \& {Zane}}]{Campana2007}
{Campana} S., {Rea} N., {Israel} G.~L., {Turolla} R., {Zane} S. 2007, \aap,
  463, 1047

\bibitem[{{Costa} {et~al}\mbox{.}(2010){Costa}, {Bellazzini}, {Tagliaferri},
  {Matt}, {Argan}, {Attin{\`a}}, {Baldini}, {Basso}, {Brez}, {Citterio}, {di
  Cosimo}, {Cotroneo}, {Fabiani}, {Feroci}, {Ferri}, {Latronico}, {Lazzarotto},
  {Minuti}, {Morelli}, {Muleri}, {Nicolini}, {Pareschi}, {di Persio},
  {Pinchera}, {Razzano}, {Reboa}, {Rubini}, {Salonico}, {Sgro'}, {Soffitta},
  {Spandre}, {Spiga}, \& {Trois}}]{Costa2010c}
{Costa} E. {et~al.} 2010, Experimental Astronomy, 28, 137

\bibitem[{{Costa} {et~al}\mbox{.}(2001){Costa}, {Soffitta}, {Bellazzini},
  {Brez}, {Lumb}, \& {Spandre}}]{Costa2001}
{Costa} E., {Soffitta} P., {Bellazzini} R., {Brez} A., {Lumb} N., {Spandre} G.
  2001, \nat, 411, 662


\bibitem[\protect\citeauthoryear{Durant \& van Kerkwijk}{2006}]{dkvk06}
Durant M., van Kerkwijk M.H. 2006, \apj, 650, 1070


\bibitem[{{Dov{\v c}iak} {et~al}\mbox{.}(2011){Dov{\v c}iak}, {Muleri},
  {Goosmann}, {Karas}, \& {Matt}}]{Dovciak2011}
{Dov{\v c}iak} M., {Muleri} F., {Goosmann} R.~W., {Karas} V., {Matt} G. 2011,
  \apj, 731, 75

\bibitem[{{Fabiani et al.}(2013)}]{Fabiani2013}
{Fabiani S. et al.}, 2013, ApJ, submitted

\bibitem[\protect\citeauthoryear{Fern\'{a}ndez \& Thompson}{2007}]{ft07}
Fern\'{a}ndez R., Thompson C. 2007, ApJ, 660, 615

\bibitem[\protect\citeauthoryear{FD}{2011}]{Fernandez+Davis}
Fern\'{a}ndez R., Davis S.W. 2011, ApJ, 730, 131

\bibitem[\protect\citeauthoryear{Goldreich \& Julian}{1969}]{Goldreich+Julian}
Goldreich P., Julian, W.H. 1969, MNRAS, 389, 157, 869

\bibitem[\protect\citeauthoryear{Harding \& Lai}{2006}]{Harding+Lai}
Harding A.K., Lai D. 2006, Rep. Prog. Phys. 69, 2631

\bibitem[\protect\citeauthoryear{Heyl \& Shaviv}{2000}]{hs00}
Heyl J.S., Shaviv N.J. 2000, MNRAS, 311, 555

\bibitem[\protect\citeauthoryear{Heyl \& Shaviv}{2002}]{hs02}
Heyl J.S., Shaviv N.J. 2002, Phys. Rev. D, 66, 023002

\bibitem[\protect\citeauthoryear{Lai et al.}{2010}]{Laietal}
Lai D., Ho W.C.G., Van Adelsberg M., Wang C. \& Heyl J.S. 2010,
X-ray Polarymetry: A New Window in Astrophysics (Cambridge: Cambridge University Press)

\bibitem[\protect\citeauthoryear{Lyutikov \& Gavriil}{2006}]{lg06}
Lyutikov M.,  Gavriil F. P. 2006, MNRAS, 368, 690

\bibitem[\protect\citeauthoryear{Mereghetti}{2008}]{mereg08}
Mereghetti S. 2008, A\&A Rev., 15, 225

\bibitem[{{Muleri} {et~al}\mbox{.}(2008){Muleri}, {Soffitta}, {Baldini},
  {Bellazzini}, {Bregeon}, {Brez}, {Costa}, {Frutti}, {Latronico}, {Minuti},
  {Negri}, {Omodei}, {Pesce-Rollins}, {Pinchera}, {Razzano}, {Rubini},
  {Sgr{\'o}}, \& {Spandre}}]{Muleri2008}
{Muleri} F. {et~al.} 2008, Nuclear Instruments and Methods in Physics Research
  A, 584, 149

\bibitem[{{Muleri} {et~al}\mbox{.}(2010){Muleri}, {Soffitta}, {Baldini},
  {Bellazzini}, {Brez}, {Costa}, {Fabiani}, {Krummenacher}, {Latronico},
  {Lazzarotto}, {Minuti}, {Pinchera}, {Rubini}, {Sgro}, \&
  {Spandre}}]{Muleri2010}
{Muleri} F. {et~al.} 2010, Nuclear Instruments and Methods in Physics Research
  A, 620, 285

\bibitem[\protect\citeauthoryear{NTZ}{2008a}]{NTZ1}
Nobili L., Turolla R., Zane S. 2008a, MNRAS, 386, 1527

\bibitem[\protect\citeauthoryear{Nobili, Turolla \& Zane}{2008b}]{ntz2}
Nobili L., Turolla R., Zane S., 2008b, MNRAS, 389, 989

\bibitem[\protect\citeauthoryear{Pavan et al.}{2009}]{pav09}
Pavan L., Turolla R., Zane S., Nobili, L. 2009, MNRAS, 395, 753

\bibitem[\protect\citeauthoryear{Perna \& Pons}{2011}]{pp11}
Perna R., Pons J.A. 2011, ApJ, 727, L51

\bibitem[{{Rea} {et~al}\mbox{.}(2005){Rea}, {Oosterbroek}, {Zane}, {Turolla},
  {M{\'e}ndez}, {Israel}, {Stella}, \& {Haberl}}]{Rea2005}
{Rea} N., {Oosterbroek} T., {Zane} S., {Turolla} R., {M{\'e}ndez} M., {Israel}
  G.~L., {Stella} L., {Haberl} F. 2005, \mnras, 361, 710

\bibitem[\protect\citeauthoryear{Rea et al.}{2008}]{reaetal08}
Rea N., Zane S., Turolla R., Lyutikov M., G\"otz D. 2008, ApJ,
686, 1245

\bibitem[\protect\citeauthoryear{Rea et al.}{2010}]{reaetal10}
Rea N. et al. 2010, Science, 330, 944

\bibitem[\protect\citeauthoryear{Rea \& Esposito}{2011}]{reaesp11}
Rea N., Esposito P. 2011, in High-Energy Emission from Pulsars and
their Systems, Astrophysics and Space Science Proceedings,
Springer-Verlag Berlin Heidelberg, p. 247

\bibitem[\protect\citeauthoryear{Rea et al.}{2012}]{reaetal12}
Rea N., Pons J.A., Torres D.F., Turolla R. 2012, ApJ, 748, L12

\bibitem[{{Soffitta et al.}(2013)}]{Soffitta2013}
{Soffitta, P. et al.} 2013, Experimental Astronomy, in press

\bibitem[{{Soffitta} {et~al}\mbox{.}(2013){Soffitta}, {Muleri}, {Fabiani}, {Costa},
  {Bellazzini}, {Brez}, {Minuti}, {Pinchera}, \&
  {Spandre}}]{Soffitta2013b}
{Soffitta} P. {et~al.} 2013, Nuclear Instruments and Methods in Physics Research
  A, 700, 99

\bibitem[{{Tagliaferri} {et~al}\mbox{.}(2012){Tagliaferri}, {Hornstrup},
  {Huovelin}, {Reglero}, {Romaine}, {Rozanska}, {Santangelo}, \&
  {Stewart}}]{Tagliaferri2012}
{Tagliaferri} G., {Hornstrup} A., {Huovelin} J., {Reglero} V., {Romaine} S.,
  {Rozanska} A., {Santangelo} A., {Stewart} G. 2012, Experimental Astronomy,
  34, 463

\bibitem[\protect\citeauthoryear{TLK}{2002}]{TLK}
Thompson C., Lyutikov M., Kulkarni, S.M. 2002, ApJ, 574, 332

\bibitem[\protect\citeauthoryear{Tiengo et al.}{2013}]{tiengoetal13}
Tiengo A., et al. 2013, \nat, 500, 312

\bibitem[\protect\citeauthoryear{Turolla \& Esposito}{2013}]{turesp13}
Turolla R., Esposito P. 2013, Int. J. Mod. Phys. D, 22, 1330024 
[arXiv:1303.6052T]

\bibitem[\protect\citeauthoryear{Van Adelsberg \& Lai}{2006}]{vanadlai06}
van Adelsberg M., Lai D. 2006, MNRAS, 373, 1495

\bibitem[\protect\citeauthoryear{Zane et al.}{2009}]{zrtn09}
Zane S., Rea N., Turolla R., Nobili L. 2009, MNRAS, 398, 1403


\end{thebibliography}
\end{document}